\documentstyle[twocolumn,aps,prb]{revtex}

\begin{document}
\draft
\title{Resistance effects due to magnetic guiding orbits}
\author{J. Reijniers and F. M. Peeters \cite{peeters}}
\address{Departement Natuurkunde, Universiteit Antwerpen (UIA),
Universiteitsplein 1, B-2610 Antwerpen, Belgium}
\date{\today}
\maketitle

\begin{abstract}
The Hall and magnetoresistance of a two dimensional electron gas
subjected to a magnetic field barrier parallel to the current
direction is studied as function of the applied perpendicular
magnetic field. The recent experimental results of Nogaret {\em
et al.} [Phys. Rev. Lett. {\bf 84}, 2231 (2000)] for the magneto-
and Hall resistance are explained using a semi-classical theory
based on the Landauer-B\"{u}ttiker formula. The observed positive
magnetoresistance peak is explained as due to a competition
between a decrease of the number of conducting channels as a
result of the growing magnetic field, from the fringe field of
the ferromagnetic stripe as it becomes magnetized, and the
disappearance of snake orbits and the subsequent appearance of
cycloidlike orbits.

\end{abstract}


\section{Introduction}
Recently, there has been a growing experimental and theoretical
activity directed towards an increased functionality of present
day electronic devices. Previously, electrical potentials were
used to modify the current, while more recently one became
interested in the effects of magnetic field profiles, modulated
or not, on the motion of electrons in semiconductor structures.
The latter is usually a heterostructure which contains a
two-dimensional electron gas (2DEG). Inhomogeneous magnetic field
profiles in the 2DEG are created by depositing superconducting or
ferromagnetic materials on top of the heterostructure which is
then patterned in the desired shape using modern
nanolithography.\cite{peeters99}

These hybrid systems are important from a theoretical and
technological point of view, since they open the door to new
physics which might result in e.g. new magneto-electronic
devices.\cite{deboeck99} An example of such a new device is the
Hybrid Hall effect device \cite{johnson97,reijniers98} in which
the magnetic material provides a local magnetic field which
influences locally the electron transport in the underlying 2DEG.
The 2DEG then acts as a detector\cite{peeters98} measuring the
magnetic state of the magnetic material.

The fringe field arising from a magnetic stripe forms a magnetic
barrier for the electron motion in the
2DEG.\cite{peeters93,matulis94,kubrak99,vancura,kato99,matulis00}
Barriers can be created in which the sign of the magnetic field
alters in different regions of space. Due to this magnetic
gradient, electrons can be bound at the boundary line between two
regions of opposite magnetic field. The spectrum and the
corresponding magnetic edge states have been studied recently.
\cite{peeters93,muller92,ibrahim95,zwerschke99,bykov00,reijniers}

When an (1D) inhomogeneous magnetic field is applied across a
quasi 1D wire, these magnetic edge states are confined
electrically due to the wire confinement potential and they mix
with the ordinary edge states.\cite{gu97} Such a situation was
recently realized by Nogaret \emph{et al.},\cite{nogaret00} where
the inhomogeneous magnetic profile was arising from a
perpendicularly magnetized ferromagnetic stripe grown on top of
the 2DEG. They measured the magneto- and Hall resistance as
function of a background magnetic field, and observed a sharp
resistance resonance effect, which they attributed to the
formation and subsequent killing of magnetic edge states.

In the present work we give a detailed theoretical analysis of
this experiment, using a semi-classical approach in which we
consider the electrical and magnetic confinement quantum
mechanically, and include scattering processes using classical
arguments. Both, the measured Hall resistance and the
magnetoresistance will be explained. We will show that the
theoretical picture of Nogaret \emph{et al.} only captures part
of the physics which is involved and is unable to predict the
correct position of the peak in the magnetoresistance and the
Hall resistance.

 The side and top view of the experimental setup of Nogaret {\it et al.} \cite{nogaret00} are shown in
Fig.~\ref{fig:setup}. A Hall device consisting of a $W=2\mu$m
wide 2DEG channel in a GaAs/AlGaAs-heterojunction was fabricated,
with electron density $n_e=1.94\times 10^{15}$m$^{-2}$ and mean
free path $\ell=4.5\mu$m at 4.5K. A narrow ($W_f=0.5\mu$m)
$32\mu$m long ferromagnetic (Fe or Ni) stripe (thickness
$d_f=200\mu$m) was grown a distance $h=80nm$ above the center of
the electron channel.

The electron transport in the 2DEG is only influenced by the
perpendicular component of the magnetic stray field. In absence of
any background magnetic field the ferromagnetic stripe is
magnetized along the easy axis, i.e. the $y$-direction, and the
fringe field is situated outside the quasi
 1D wire, i.e. in reservoir 1 and 2. Application of a
perpendicular background magnetic field rotates the magnetization
to align with the $z$-axis, and this will result in a stray field
in the wire, which imposes a step magnetic field profile along
the $y$-direction (see Fig.~\ref{fig:setup}(b)). The actual
magnetic field profile is slightly rounded (see
Ref.\onlinecite{nogaret00}) but we checked that our results are
not influenced by this simplification.  This magnetic step adds an
inhomogeneous magnetic field component to the uniform applied
magnetic field $B_{a}$ which induces the observed resistance
effects. In the present analysis, we restrict ourselves to a
Fe-stripe (saturation magnetization: 1.74 T), since this was
studied most thoroughly in Ref.~\onlinecite{nogaret00} and
produced the most pronounced resonance effect.

This paper is organized as follows. In Sec.~II we present our
theoretical approach. In Sect.~III we calculate the two terminal
resistance as function of the applied background magnetic field.
The Hall resistance is studied in Sect.~IV and in Sect.~V the
magnetoresistance is calculated. We will discuss differences
between our theoretical results and the experimental (and
theoretical) results of Nogaret \emph{et al.}\cite{nogaret00} Our
theoretical explanation for the observed resonance effect in the
magnetoresistance deviates from the one proposed in
Ref.~\onlinecite{nogaret00}. In Sect.~VI we summarize our
conclusions.

\section{Theoretical approach}
The magneto- and Hall resistance are measured experimentally by
use of a {\em four-terminal} configuration. In contrast to the
theoretical study of Nogaret {\it et al.}\cite{nogaret00}, we
will retain this feature in the present discussion. The
four-terminal configuration is schematically shown in
Fig.~\ref{fig:four-terminal} for (a) a Hall measurement and (b) a
magnetoresistance measurement. The leads are in thermodynamical
equilibrium and can be characterized by a chemical potential
$\mu_{i}$. Each reservoir injects a current $I_{i}$ of electrons
into the 1D wire. If several bands are occupied, we have to
consider a many-channel situation, and according to B\"{u}ttiker,
\cite{buttiker86} the current in each of the leads is given by
\begin{equation}
I_{i}=\frac{e}{h}\sum_{n,n'}\left\{
[\delta_{n,n'}-T_{ii}(n,n')]\mu _{i} -\sum_{j\neq i}T_{ij}(n,n')
\mu _{j}\right\}, \label{eq:lan-but}
\end{equation}
where $T_{ij}(n,n') $ is the probability for an electron in
channel $n$ of lead $i$ to be scattered/transmitted to $n'$ of
lead $j$. Current conservation requires $N_{i}=R_{ii}+\sum_{j\neq
i}T_{ij}$ for all $i$, with $T_{ij}=\sum_{n,n'}T_{ij}(n,n')$ and
$R_{ii}=1-T_{ii}$ and $N_{i}$ is the number of channels in lead
$i$.

Each channel $n$ contributes a probability $T_{ij}(n)$ to the
conductivity which is transmitted from probe 1 to probe 2. The
total transmission from probe $i$ to $j$ then equals
$T_{ij}=\sum_{n \le N}T_{ij}(n)$, and Eq.~(\ref{eq:lan-but}) is
simplified to
\begin{equation}
 I_{i}=\frac{e}{h}\sum_{n}\left\{
[1-R_{ii}(n)]\mu _{i} -\sum_{j\neq i}T_{ij}(n) \mu _{j}\right\}.
\end{equation}

In this type of measurement, only two probes are current
carrying, i.e., $i=1,2$, which results in the condition
$I_{1}=-I_{2}=I$ while the other probes are voltage probes and do
not carry any net current: $I_{3}=I_{4}=I_{5}=0$.

In order to calculate the four-terminal magneto- and Hall
resistance, we will make another simplifying assumption that the
voltage probes are weakly coupled, i.e. their influence on the
net current $I$ is very small ($I=(\mu_1-\mu_2)/R_{12,12}$) and
the chemical potentials in each of the voltage probes can be
calculated in the absence of the other voltage probes
($\mu_i=(T_{i1}\mu_1+T_{i2}\mu_2)/(T_{i1}+T_{i2})$ with $i=3,4$).
The general formula for this kind of resistance measurement is
then readily obtained and given by
\begin{eqnarray*}
R_{12,3i}
&=&\frac{\mu_{3}-\mu_{i}}{eI}=\frac{h}{e^{2}}\frac{1}{T_{12}}
\frac{T_{31}T_{i2}-T_{32}T_{i1}}{( T_{31}+T_{32})(T_{i1}+T_{i2})}\\
&=&\frac{h}{e^{2}}\frac{1}{T_{12}}F=R_{12,12}F,
\end{eqnarray*}
which is the two terminal resistance $R_{12,12}$ multiplied with
a geometrical form factor $F$, which is less than one.

In the following we will first calculate the two-terminal
resistance $R_{12,12}$ and then concentrate on the geometrical
form factor $F$ in the case of a Hall or magnetoresistance
measurement.

\section{The energy spectrum and the two-terminal resistance}
The two-terminal resistance is given by
$R_{12,12}=(\mu_2-\mu_1)/I$. We know that in the absence of any
collisions, the current which flows from reservoir 1 to 2 is
determined by the number of subbands $N$ which are occupied at
the Fermi level. Since the mean free path in the experiment of
Nogaret {\em et al.}\cite{nogaret00} is $\ell=4\mu $m, which is
larger than the wire width ($W=2\mu m$), we can, to a good
approximation, neglect the influence of scatterers on the spectrum
and calculate the number of channels quantum mechanically
following the work of M\"{u}ller\cite{muller92} for a pure quasi
1D quantum wire.

We consider a system of noninteracting electrons moving in the
$xy$-plane subjected to a hard wall confinement $-W/2<x<W/2$,
where $W$ is the width of the wire. The electrons are subjected
to a magnetic field profile $\overrightarrow{B}=(0,0,B_z(x))$.
This profile equals $B_z=B_i(B_a)+B_a$, where $B_a$ is the
uniform applied background field and $B_i$ is the induced magnetic
field profile due to the magnetized stripe.

In correspondence with Ref.~\onlinecite{nogaret00} we will model
the shape of the induced magnetic field profile by the average
magnetic field on the respective sides of the magnetic stripe
edges, i.e. at saturation the magnetic field profile is given by
$B_{sat}=B_1-(B_1+B_2)\theta(|x|-W_f/2)$, where $\theta$ is the
heavyside step function and $B_1=-0.06$ Tesla and $B_2=0.28$
Tesla are the modeled magnetic field strengths underneath and
away from the stripe as shown in Fig.~\ref{fig:setup}(c). We also
performed the calculations for the exact magnetic field profile,
but this resulted in negligible small quantitative differences.

We model the magnetization of the stripe by considering two
limiting cases: (A) when the stripe is already magnetized at
$B_a=0$ Tesla (as was considered by Nogaret \emph{et al}.), i.e.
$B_i=B_{sat}sign(B_a)$ which is the hard magnet case, and (B)
when the applied magnetic field magnetizes the stripe as for soft
magnets. In case (B) we assume $B_i$ to be linearly varying with
applied background magnetic field $B_a$, up to $B_a=0.05$ Tesla,
where saturation is attained according to
Ref.~\onlinecite{nogaret00}. The induced magnetic field is then
given by $B_i=B_{sat}\{1-[1-\theta(|B_a|-0.05)](1-|B_a|/0.05)\}$.
The actual experimental behaviour is expected to be situated
closer to situation (B) than to (A).

The one-particle states are described by the Hamiltonian
\begin{equation}
 H=\frac{1}{2m_e}p^2_x+\frac{1}{2m_e}\left[
p_y-\frac{e}{c}A(x) \right]^2+V(x),
\end{equation}
where $V(-W/2<x<W/2)=0$ and $V(x<-W/2)=V(x>W/2)=\infty$. Taking
the vector potential in the Landau gauge
$\overrightarrow{A}=(0,A_y(x),0)$, such that $\partial
A_y(x)/\partial x= B_z(x)$, we arrive at the following 2D
Schr\"{o}dinger equation
\begin{equation}
\left\{ \frac{\partial^2}{\partial
x^2}+\left[\frac{\partial}{\partial y} + A_y(x)
\right]^2+2[E-V(x)] \right\}\psi(x,y)=0,
\end{equation}
where the magnetic field is expressed in $B_0$, magnetic units
are used for a homogeneous field of $B_0=1$ Tesla, i.e., all
lengths are measured in $\l_0=\sqrt{\hbar c/eB_0}=0.0257\mu m$
and energy is measured in units of $E_0=\hbar eB_0 /m_ec=1.7279
meV$. $H$ and $p_y$ commute due to the particular choice of the
gauge, and consequently the wavefunction becomes
\begin{equation}
\psi(x,y)=\frac{1}{\sqrt{(2\pi)}}e^{-iky}\phi_{n,k}(x),
\end{equation}
which reduces the problem to the solution of the 1D
Schr\"{o}dinger equation
\begin{equation}
\left[ -\frac{1}{2}
\frac{d^2}{dx^2}+V_k(x)\right]\phi_{n,k}(x)=E_{n,k}\phi_{n,k}(x),
\label{eq:1Dschrod}
\end{equation}
where it is the $k$-dependent effective potential
\begin{equation}
V_k(x)=\frac{1}{2}[xB_z(x)+k]^2+V(x),
\end{equation}
which contains the two dimensionality of the
problem.\cite{peeters93} We solve Eq.~(\ref{eq:1Dschrod})
numerically by use of a discretization procedure.

For given applied background magnetic field we calculated the
energy spectrum for case (A) with $W=2\mu m$. The results are
shown in Fig.~\ref{fig:spectra1} for $B_a/B_0=0;\,0.1;\,0.5$.
These energy spectra are symmetric in $k$ and for small $B_a$
consist of the superposition of two parabolic spectra. For small
$k$-values and for energies below the intersection of the two
parabolas Landau levels are present due to electrons which are
bound underneath the stripe. These levels shift away from each
other as the background magnetic field increases, due to the
increase of the magnetic field underneath the stripe ($B_1$). For
increasing magnetic field the two parabolas shift further away
from each other, towards higher $|k|$ values. Due to the
confinement of the wire, each parabola is infinitely dublicated,
where its maximum is shifted to higher energy and to lower
$k$-values. For higher magnetic fields ($B_a>0.5B_0$) Landau
levels arise, due to the magnetic field away from the stripe
($B_2<B_1$) which is now strong enough to localize electrons into
cyclotron orbits.

The classical trajectories (for $E=4E_0$) corresponding to the
different regions in $k$-space are shown on top of
Fig.~\ref{fig:spectra1}. We restricted ourselves to trajectories
of states at energy $E=4E_0$, since at zero temperature only
channels with energy $E=\epsilon _{F}=4E_0=6.9\;meV$ contribute to
the conductivity.

For (B) the spectrum at $B_a/B_0=0;\,0.03;\,0.06$ and the
corresponding classical orbits at the Fermi energy are shown in
Fig.~\ref{fig:spectra2}. For $B_a=0$ the magnetic stripe is not
magnetized and the spectrum consists only of the potential
confined levels. One single parabola (and its duplicates due to
confinement) centered around $k=0$ is found which splits into two
and its center shifts towards higher $|k|$-values. Below the
intersection of the two shifted parabolas Landau states are
formed. Notice that some levels intersect the Fermi energy twice
as much as before. These Landau states separate further away from
each other for increasing magnetic field. For $B_a \geq 0.05B_0$
the spectra are identical to the ones of (A).

The number of conducting channels is given by the number of
energy levels intersecting the Fermi energy $N$, and hence the
current is
\begin{equation}
I =\frac{e}{h}T_{12}( \mu _{i}-\mu _{j} )
\end{equation}
with $T_{12}=\sum_{n<N}T_{12}(n)=N$, since in absence of any
collisions $T_{12}(n)=1$.

Nevertheless, the mean free path measured by Nogaret {\em et al.}
is smaller than the length of the wire $L_{y}=16\;\mu m$ and also
smaller than the distance between the probes. Thus, scattering
will play an important role in electron transport and consequently
$T_{12}(n)$ will be less than 1. In order to account for this, we
will estimate the transmission coefficient for every channel using
classical arguments. Since we consider the voltage probes as
weakly coupled, they result in a very weak perturbation of the
electron-current path, and scattering due to the voltage probes
will be neglected. The only scattering we consider is due to
collisions with impurities and other imperfections in the
1D-channel.

The rate at which these collisions occur depends classically on
the velocity in the $y$-direction, the length of the wire and the
scattering time. The lower the velocity in the $y$-direction, the
longer it takes to overcome the distance between probe 1 and
probe 2, and the more probable it will be to experience a
scattering event. Because of this we consider the transmission
probability of every channel to be proportional to its velocity
$v_y=-\partial E_{n}(k)/\partial k |\epsilon_F$ and the
scattering time $\tau$, and inversely proportional to the length
$L_y$ of the wire, i.e.,
\begin{equation}
T_{12}(n)\sim \frac{v_{y}(n)\tau}{L_{y}}.
\end{equation}
So finally, we arrive at the two terminal resistance
\begin{equation}
R_{12,12}=\frac{1}{\alpha}\frac{\hbar}{e^2}\frac{1}{\sum_{n}v_y(n)},
\end{equation}
where $n$ runs over all the $N$ electron states with positive
velocity (or negative velocity) at the Fermi energy $\epsilon_F$,
and $\alpha$ is a function of $L_y$ and $\tau$.

First we will discuss the change of the two-terminal resistance
$R_{12,12}$ with respect to the situation in absence of the
ferromagnetic stripe $R_{12,12}^0$, which we will call the
\emph{induced} resistance $R_{12,12}/R_{12,12}^0$. This property
was also calculated and discussed by Nogaret \emph{et
al.},\cite{nogaret00} and is plotted in
Fig.~\ref{fig:induced-res} as function of the applied magnetic
field $B_a$ for the approach of Ref.~\onlinecite{nogaret00}
(dashed curve) and ours (dotted and solid curves). The zero
temperature result is shown in the inset.

We notice that at zero temperature, many discontinuous jumps are
present. As we will see further on, their position is very
sensitive to the Fermi-energy and they disappear at 4.2 K. The
energy distribution function
$f(E,T)=\{\exp[(E-\epsilon_F)/k_BT]+1\}^{-1}$ is not a
stepfunction for nonzero temperature and consequently also
electrons with energy different from $\epsilon_F$ will contribute
to the conductivity. This will smoothen out these oscillations,
as is shown in Fig.~\ref{fig:induced-res}. Also the broadening of
the energy levels due to e.g. potential fluctuations will have
such a smoothening effect on the resistance curves. Hence, we will
only show these smooth curves in the next figures.

The curves (A) and (B) differ only for $B_a<0.05B_0$. In case (A)
the resistance for $B_a=0$ is larger than in the absence of the
magnetic stripe. Increasing the background magnetic field results
in a slight overall increase of the induced resistance. At
$B_a=0.02B_0$ the induced resistance reaches its maximum, then it
decreases rapidly.

The induced resistance in case (B) starts at 1 for $B_a=0$,
increases more rapidly and attains its maximum at a slightly
higher $B_a$-value, i.e., $B_a=0.0375B_0$. Then it decreases
rapidly up to $B_a=0.2B_0$. We again notice oscillations at zero
temperature (see inset), but fewer than for case (A). For larger
$B_a$-values the oscillations disappear and the scaled resistance
increases ultimately to one.

Nogaret \emph{et al.}\cite{nogaret00} obtained theoretically a
somewhat similar behaviour, as is indicated by the dashed curve,
except for the peak which was situated at a slightly higher value
$B_a=0.06B_0$. They made the assumption that the stripe was
already fully magnetized at $B_a=0$ like in our case (A).
Moreover they considered the magnetoresistance for a homogeneous
magnetic field profile with magnetic field strength $B_a$, and
considered the effect of the magnetic stripe profile by adding
classical trajectories of states which arise due to the presence
of the stripe. In order to simplify the problem, they only
considered states which do not reach (classically) the edge of
the sample. They attributed the initial positive
magnetoresistance to \emph{snake orbits} (see situation ``$\circ$"
in Fig.~\ref{fig:spectra1}(a)) which are killed with increasing
magnetic field and therefore no longer contribute to the
conductivity for larger fields. At $B_a=0.06B_0$ all snake orbits
have vanished and it is due to this, they inferred, that the
resistance reaches its maximum. However if the magnetic field is
larger than $0.06B_0$, the magnetic field has the same sign over
the whole sample but has different strength under the stripe and
away from it, and a new type of magnetic edge states, so called
\emph{cycloidlike} states (see the states indicated by
``$\triangleleft$" in Fig.~\ref{fig:spectra1}(b)), arises, which
again enhances conductivity and thus lowers the resistance. The
fact that the influence of the latter orbits vanishes for larger
$B_a$-values is due to the decrease of the velocity of these
states with decreasing relative difference between the two
neighboring magnetic fields.

In case (A), when the saturation magnetization is already attained
at $B_a=0$, we cannot attribute the existence of the (small) peak
to the creation or annihilation of a certain classical state.
Fig.~\ref{fig:spectra1} shows the energy spectrum and the
corresponding classical states at the Fermi-energy
$\epsilon_F=4E_0$ for (a) $B_a=0$, (b) $B_a=0.1B_0$ and (c)
$B_a=0.5B_0$. From this figure we see that the enhancement of the
resistance is a pure quantum mechanical effect and involves many
different types of states with different velocities. Therefore an
explanation based on the appearance or disappearance of only
snake orbits as done by Nogaret \emph{et al.} is not possible, at
least for small $B_a$. The discontinuous behaviour for small
$B_a$ (see the inset in Fig.~\ref{fig:induced-res}) is due to edge
states at the Fermi-level whose energy moves through the Fermi
level and then no longer contribute to the conduction. They have
nonzero velocity and hence this is also reflected in the
resistance. For larger $B_a>0.05B_0$ the curve coincides with the
one of case (B).

In case (B), the initial magnetoresistance can be understood more
easily. For $B_a=0$, the stripe is not yet magnetized and thus
there is no effect of the magnetic stripe.
Fig.~\ref{fig:spectra2}(a) shows subbands formed due to the quasi
1D confinement ($N=70$ subbands contribute to the conduction).
Already for a small applied magnetic field, a relative large
magnetic field is induced in the wire due to the magnetization of
the ferromagnetic stripe. Whereas, for $B_a=0$ the only
confinement was due to the edges of the sample, the magnetic
field tends to localize electrons into cyclotron orbits and thus
forces them in Landau levels, which separate further in energy
with increasing magnetic field. As a consequence less channels
will intersect the Fermi-level and consequently less channels
contribute to electron transport and the resistance increases.

But there is a competing effect due to the presence of the
magnetic stripe which tends to lower the induced resistance: new
edge states arise (see the states indicated by ``$\triangleleft$"
and ``$\triangleright$" in Figs.~\ref{fig:spectra1} and
\ref{fig:spectra2}) which travel in the opposite direction of the
normal edge states. In terms of energy spectra, this effect is
contained in subbands which contribute twice to the conductivity,
as can be seen in Figs.~\ref{fig:spectra1}(b,c). It overcomes the
previous one at $B_a=0.0375B_0$, and the induced resistance
decreases. This effect contributes even for $B_a>0.06B_0$, when
the magnetic field has the same sign in the whole wire, and the
previously mentioned cycloidlike orbits appear. For increasing
magnetic field their influence decreases, although their number
with respect to normal edge states increases. This is due to
their velocity, which decreases for increasing $B_a$ for reasons
given by Nogaret \emph{et al.}


In the following sections, we will try to reproduce the
experimental results obtained by Nogaret \emph{et al.} First we
will concentrate on the Hall resistance, before we focus on the
magnetoresistance.

\section{The Hall resistance}
In order to calculate the magneto- and Hall resistance we will
further simplify the problem, by making the assumption of
symmetrical probes. In case of the Hall resistance, the two
voltage probes, i.e. probe 3 and 4, are in front of each other as
is clear from Fig.~\ref{fig:four-terminal}, and due to this
symmetry the transmission probabilities can be written as
$T_{31}=T_{42}=T$ and $T_{32}=T_{41}=t$. The Hall resistance then
attains a very simple form
\begin{equation}
R_{12,34}=\frac{1}{\alpha}\frac{h}{e^{2}}\frac{1}{T_{12}}\frac{\left(
T/t-1\right) }{\left( T/t+1\right) },
\end{equation}
as was already derived in Ref.~\onlinecite{peeters88}. Note that
there is additionally one parameter $\alpha$ and one function
$t/T$ which describe the behaviour of the Hall resistance. In the
absence of any magnetic field $t/T=1$, and consequently the Hall
resistance reduces to zero. As the cyclotron radius decreases,
electrons will be localized closer to the edge (in edge states)
and consequently the probability for an electron
\textit{bouncing} on one edge to be transmitted in a probe on the
other side of the wire decreases drastically, i.e. exponentially,
with increasing magnetic field, as can be inferred from
Ref.\onlinecite{peeters88}. Consequently $t/T$ will decrease
rapidly and ultimately for already a small applied magnetic field
the geometrical form factor $F=\left( T/t-1\right) /\left(
T/t+1\right) $ will be 1 in which case the Hall resistance equals
the two-terminal resistance $R_{12,12}$.

In order to obtain qualitative agreement with experiment, we
follow Ref.~\onlinecite{peeters89} and take the following
functional form for $t/T = \exp[-25B_a-(35B_a)^2]$ with $B_a$
expressed in Tesla. If we take
$R_{12,34}^0(B_a)=(3669.4*B_a)\Omega$ as the functional form of
the Hall resistance in absence of the magnetic stripe, which we
obtained from a linear fit of the experimental result by Nogaret
\emph{et al.}, we obtain the induced Hall resistance
$R_{12,34}/R^0_{12,34}$ as shown in
Fig.~\ref{fig:induced-Hall-res} for case (A) and (B) (dotted and
solid curves, respectively) which is compared with the
experimental result (dashed curve).

We notice that the induced Hall resistance for $B_a=0$ approaches
0.5 in both cases (A) and (B), and increases rapidly until
$B_a=0.0325B_0$ in case (A) and $B_a=0.0375B_0$ for (B). The
experimental peak position of the Hall resistance $B_a=0.04B_0$ is
very close to these values. For larger $B_a$ the curve almost
coincides with the one in Fig.~\ref{fig:induced-res}, which is
due to the exponential form of $t/T$.

Notice that for case (B) the peak is very close to the
experimental position and the qualitative behaviour of the Hall
resistance is reproduced, the experimental curve differs
quantitatively with the theoretical one only for $B_a>0.04B_0$.
Our Hall resistance in this magnetic field range is smaller than
measured experimentally. From this comparison it seems that the
cycloidlike electron trajectories do not contribute much for large
applied magnetic field $B_a$. This might be due to a large
concentration of scatterers underneath the magnetic stripe edge,
which might arise from the fabrication process of the sample.
This would not only result in an increase of the resistance, but
would especially hamper/kill the cycloidlike states propagating
underneath it.

Due to the fact that the Hall resistance for large $B_a$ equals
the two terminal resistance, it is possible to estimate $\alpha$
by comparison of the theoretical curve with the experimental one.
In order to obtain reasonable agreement with the experimental
curve, we have to assume $\alpha=1.59$. This value is now fixed
and will be used to compare our theoretical results on the
magnetoresistance with the experimental results of
Ref.~\onlinecite{nogaret00}.

\section{The magnetoresistance}
In order to measure the magnetoresistance, the voltage probes are
on the same side of the wire and separated a distance from each
other along the 1D wire as shown schematically in
Fig~\ref{fig:four-terminal}(b). If the probes are situated on the
same side (which is the case for the curves under study), we can
approximate the transmission probabilities by: $T_{31}=T$,
$T_{32}=(1-\beta )t$, $T_{42}=(1-\beta )T$, $T_{43}=t$, where
$\beta <1$ is defined as the fraction of the current $I_{12}$
which is reflected due to collisions between probe $3$ and $4$.
The magnetoresistance is then given by
\begin{equation}
R_{12,34}=\frac{h}{e^{2}}\frac{1}{T_{12}}\frac{\left( 2-\beta \right) \beta
}{\left[ \frac{t}{T}+\frac{T}{t}\right] \left( 1-\beta \right) +\beta
^{2}-2\beta +2}.
\end{equation}
Note that in this case there are two parameters $\alpha$ and
$\beta$ and one function $t/T$ which determine the Hall
resistance.

In absence of any magnetic field, $t/T=1$ and we obtain for the
form factor $F\approx \beta/(2-\beta)$. For increasing magnetic
field $B_a$, $t/T$ will decrease rapidly and for $B_a \gg 0$,
$t/T<1$, which results in $F=(t/T)(2\beta)/(1-\beta)$.

Theoretically $\alpha$ and $t/T$ are identical to those of the
previous section and we have only to determine the parameter
$\beta$. It is clear that also this parameter depends on the
distance between the two voltage probes, the scattering time and
the velocity of the electron states at the Fermi-level. For
simplicity we will consider $\beta(B_a)\approx \beta$ to be
independent of the magnetic field, which is justified since the
function $t/T(B_a)$ changes more drastically than $\beta(B_a)$.
We will show that this approximation already results in good
qualitative agreement with the experimental curves.

If we insert $\alpha=1.59$ and $t/T$ identical to the ones
obtained from the Hall resistance, we arrive at the
magnetoresistance shown in Fig.~\ref{fig:magnetores}. In this
figure we took $\beta=0.95$ and plotted $R_{12,35}$ for case (A)
(dotted curve) and case (B) (solid curve) together with the
experimental result of Nogaret \emph{et al.} (dashed curve), which
is plotted with respect to the right hand axis. We find a peak in
the resistance at $B_a=0.02B_0$ in case (A), and $B_a=0.0275B_0$
for (B). The experimental peak position ($B_a=0.03B_0$) (dotted
curve) is very close to our theoretical result for case (B).
Notice that the peak position occurs for smaller $B_a$ then for
the induced Hall resistance, which is in correspondence with the
experimental results.

In contrast to the experimental results, we notice that for large
$B_a$-values the magnetoresistance is zero. This is due to the
fact that we have assumed that for large $B_a$-values, $t/T=0$.
But due to scattering there is always a possibility for an
electron to be scattered from an edge state localized on one side
of the sample to an edge state on the other side (and traveling
in the other direction). If we assume that this effect results in
a constant remainder $t_0=0.005$ and corresponding
$t/T=(1-t_0)\exp [-25B_a-(35B_a)^2]+t_0$, we obtain the positive
magnetoresistance as measured experimentally. This background
does not change the Hall resistance qualitatively: the slope
decreases, but this can easily be compensated with a larger
$\alpha$ in order to have good agreement with the experiment.

It is very hard to reproduce the experimental results
quantitatively, as is obvious from the need for a different left
and right axis. The magnitude of the experimental result is
larger, and an additional background is present. Due to the
approximations made in our simple approach, we underestimated the
magnetoresistance.  Moreover, the experiment also suffers from
other effects, like backscattering etc., which also influence the
resistance but which we did not take into account in this paper.
Nevertheless, we were able to reproduce the position and the
magnitude ($\approx 150\;\Omega$) of the peak.

\section{Conclusions}
In this paper we studied electron transport in a quantum wire
subjected to an abrupt magnetic field gradient arising from a
ferromagnetic stripe fabricated at its surface, as was
investigated experimentally by Nogaret {\it et
al.}\cite{nogaret00} We were able to reproduce the main
qualitative features of the magnetic field dependence of the Hall
and magnetoresistance. In particular, the position of the peak in
both resistances was correctly explained. This peak is due to two
competing effects, i.e. the increase of the separation between
subbands for increasing magnetic field, which decreases the
number of conducting channels, and the killing of snake orbits and
the creation of states which travel in the opposite direction of
ordinary snake orbits, the so called cycloidlike
states.\cite{reijniers} Two models for the magnetization of the
ferromagnetic stripe were considered corresponding to the extreme
cases of a hard (A) and a soft (B) magnet. Model (B) gives the
closest agreement with experiment which agrees with the
observation by Nogaret \emph{et al.} that almost no hysteresis
was observed.

In comparison with the theoretical approach of Nogaret \emph{et
al.},\cite{nogaret00} ours differs essentially in two ways: (1)
the magnetic field profile is the one created by a soft magnet
while Nogaret \emph{et al.} assumed a magnetic barrier which is
already present for zero applied magnetic field, and (2) we
calculated the Hall and magnetoresistance for a four probe
measurement with particular geometry, by use of a semi-classical
theory based on the Landauer-B\"{u}ttiker formula, while Nogaret
\emph{et al.}\cite{nogaret00} made use of a semi-classical drift
diffusion model.

\acknowledgments{This work was partially supported by the
Inter-university Micro-Electronics Center (IMEC, Leuven), the
Flemish Science Foundation (FWO-Vl), the ``Onderzoeksraad van de
Universiteit Antwerpen", and the IUAP-IV. J.R. was supported by
``het Vlaams Instituut voor de bevordering van het
Wetenschappelijk \& Technologisch Onderzoek in de Industrie"
(IWT) and F. M. P. is a research director with the FWO-Vl.  We
acknowledge fruitful correspondence with A. Nogaret.}

\begin{figure}[tbp]
\caption{The top (a) and side view (b) of the sample
configuration used by Nogaret \emph{et al.}\cite{nogaret00} In (c)
the resulting (modeled) magnetic field profile in the wire is
shown with $B_i$ the magnetic field profile due to the fringe
fields and $B_a$ the uniform externally applied field.}
 \label{fig:setup}
\end{figure}

\begin{figure}[tbp]
\caption{Four-terminal configuration in (a) a Hall and (b) a
magnetoresistance measurement. In a magnetic field the electron
current flows along the edge. This current is schematically shown
together with the different transmission probabilities.}
\label{fig:four-terminal}
\end{figure}

\begin{figure}[tbp]
\caption{ The energy spectrum in case (A) as function of $k$ for
(a) $B_a/B_0=0$, (b) $0.1$, and (c) $0.5$. The classical
trajectories for $\epsilon_F=4E_0$ are schematically shown on top
of the figures for the $k$-range indicated by the solid bars. The
darker area in these insets correspond to the position of the
magnetic stripe.}
 \label{fig:spectra1}
\end{figure}

\begin{figure}[tbp]
\caption{ The same as in Fig.~\ref{fig:spectra1}, but now for case
(B) for (a) $B_a/B_0=0$, (b) $0.03 $  and (c) $0.06$.}
 \label{fig:spectra2}
\end{figure}

\begin{figure}[tbp]
\caption{ The induced two point resistance
$R_{12,12}/R_{12,12}^0$ as function of $B_a$ for case (A) (dotted
curve) and case (B) (solid curve), and according to the approach
of Nogaret \emph{et al.} (dashed curve). The inset shows the zero
temperature induced resistance.}
 \label{fig:induced-res}
\end{figure}

\begin{figure}[tbp]
\caption{The induced Hall resistance $R_{12,34}/R^0_{12,34}$ as
function of the applied magnetic field $B_a$ as measured
experimentally by Nogaret \emph{et al.} (dashed curve) and our
theoretical result for case (A) (dotted curve) and case (B)
(solid curve).}
 \label{fig:induced-Hall-res}
\end{figure}

\begin{figure}[tbp]
\caption{The magnetoresistance $R_{12,35}$ as function of the
applied magnetic field $B_a$ in case (A) (dotted curve) and (B)
(solid curve). The latter is plotted with and without remainder
$t_0$. The experimental result of Nogaret \emph{et al.} (dashed
curve) is plotted with respect to the axis on the right hand side}
 \label{fig:magnetores}
\end{figure}

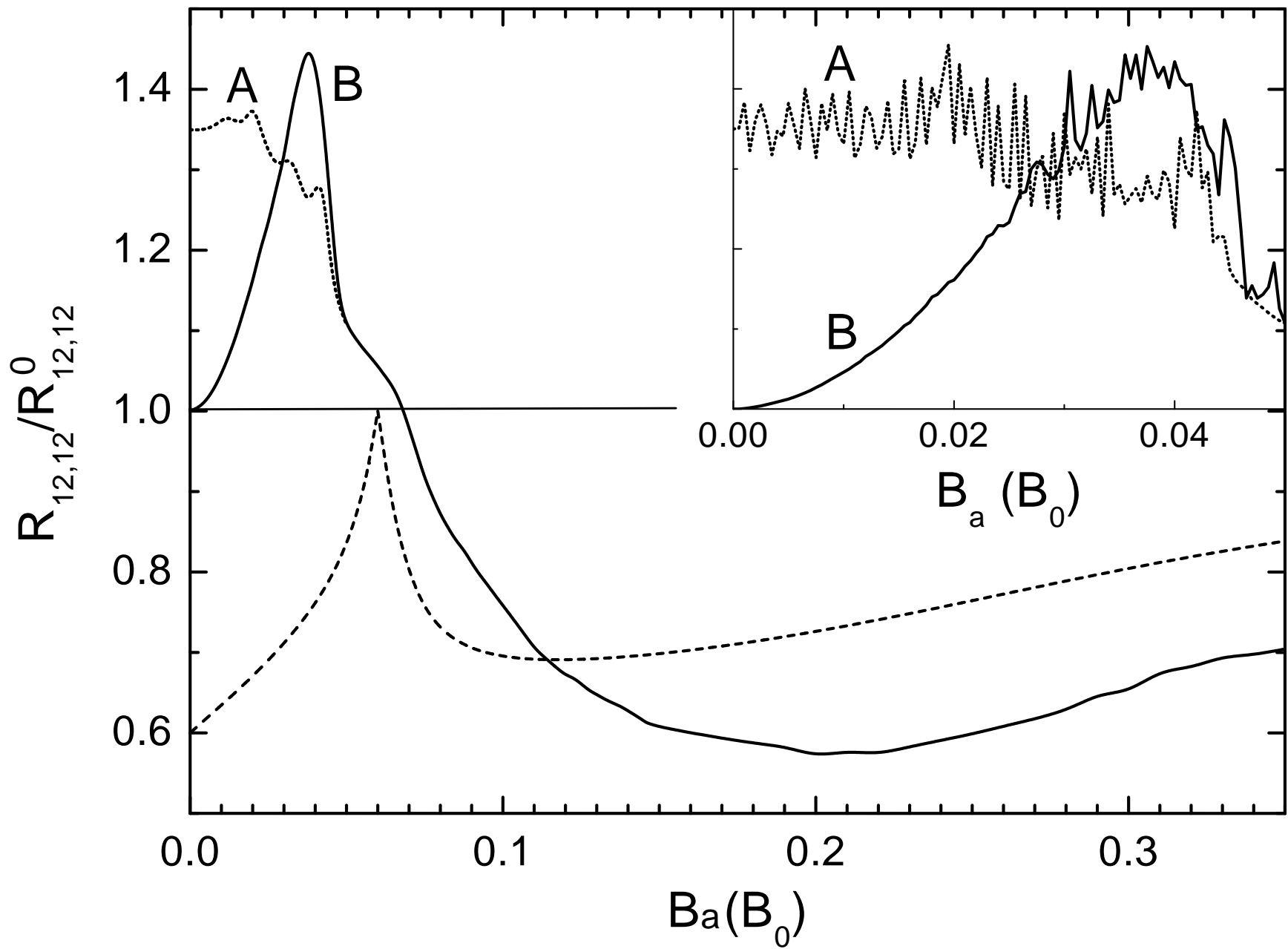
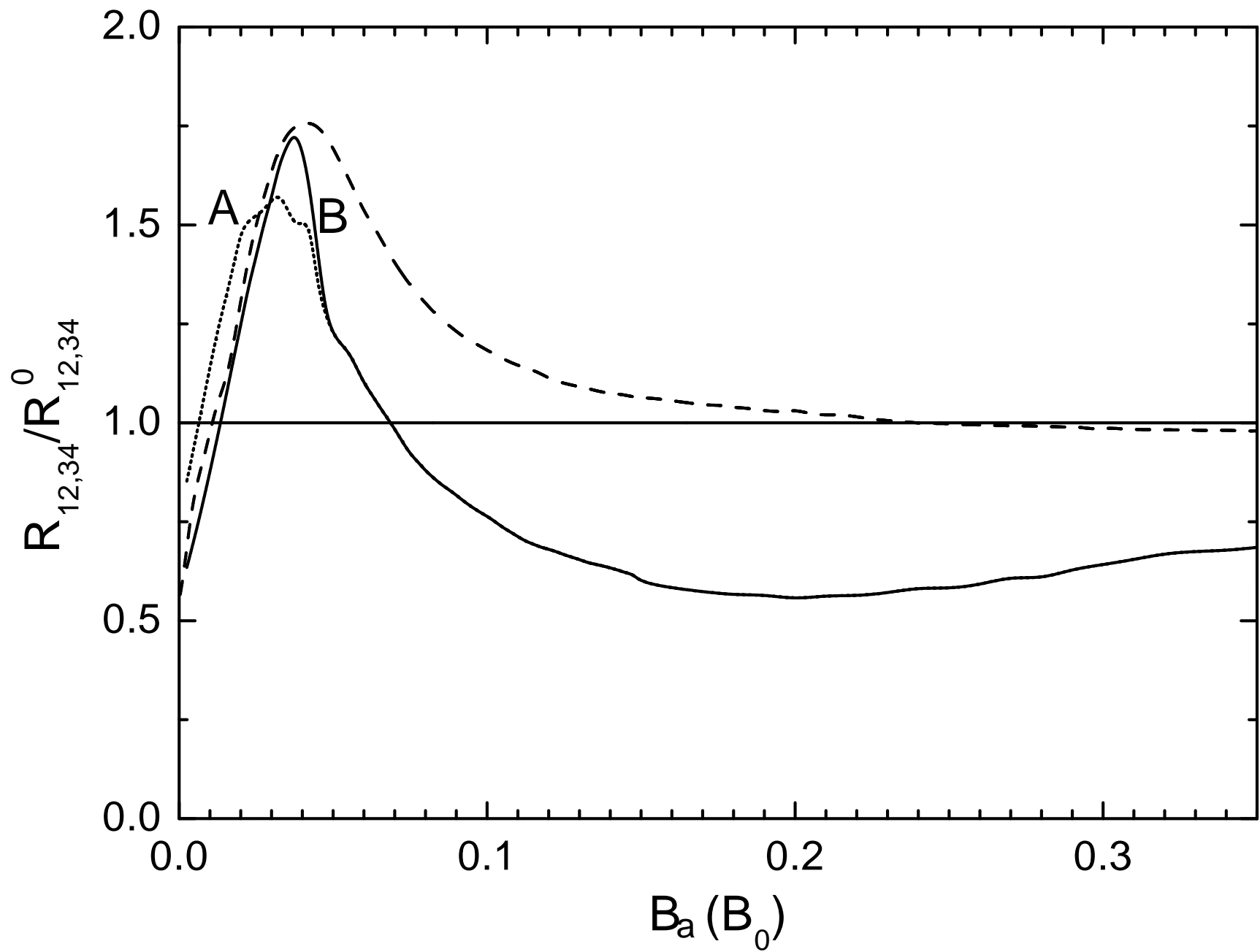
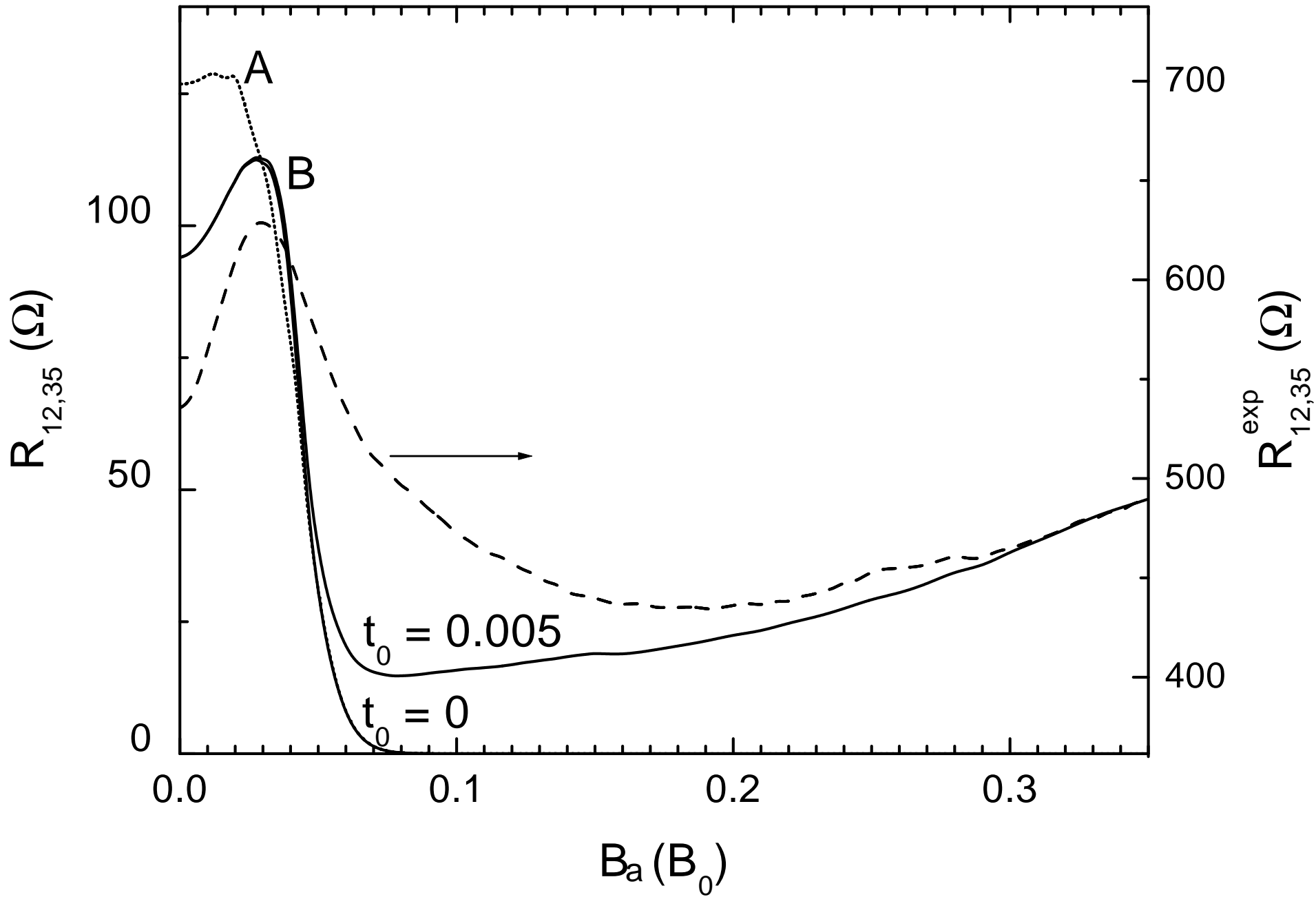
\end{document}